\documentclass[aps,pra,showpacs,twocolumn,superscriptaddress]{revtex4}
\usepackage{amsmath,amsthm,graphicx,epsfig,amssymb}

\newcommand{\tr}{\text{tr}}
\newcommand{\ten}{\otimes}
\newcommand{\ket}[1]{|#1\rangle}
\newcommand{\bra}[1]{\langle#1|}
\newcommand{\proj}[1]{\ket{#1}\!\bra{#1}}
\newcommand{\unit}{\mathbb{I}}
\newcommand{\half}{\frac{1}{2}}

\def\H{\mathcal{H}}
\def\A{\mathcal{A}}
\def\Ai{\A_{\rm in}}
\def\Ao{\A_{\rm out}}
\def\B{\mathcal{B}}
\def\Bi{\B_{\rm in}}
\def\Bo{\B_{\rm out}}
\def\E{\mathcal{E}}
\def\F{\mathcal{F}}
\def\O{\mathcal{O}}
\def\T{\mathcal{T}}
\def\HA{\mathcal{H}_\mathcal{A}}
\def\HB{\mathcal{H}_\mathcal{B}}
\def\P{\mathcal{P}}
\def\Z{\mathcal{Z}}
\def\t{^{\mbox{\tiny T}}}
\newcommand{\ii}{{\rm i}}
\newcommand{\rbd}{\rho_\text{BD}}
\newcommand{\vl}{\vec{\lambda}}
\newcommand{\ld}{\lambda}
\newcommand{\ldp}{\lambda'}
\newcommand{\vlp}{\vec{\ldp}}
\newcommand{\TE}{\T_E}

\newtheorem{theorem}{Theorem}
\newtheorem{lemma}[theorem]{Lemma}
\newtheorem{dfn}[theorem]{Definition}

\begin{document}

\title{SLOCC Convertibility between Two-Qubit States}

\author{Yeong-Cherng~Liang}\email{yeongcherng.liang@gmail.com}
\affiliation{School of Physical Sciences, The
  University of Queensland, Queensland 4072, Australia.}

\author{Llu\'{\i}s~Masanes}\email{masanes@damtp.cam.ac.uk}
\affiliation{Department of Applied Mathematics and Theoretical
Physics, University of Cambridge, Wilberforce Road, Cambridge
CB3 0WA, United Kingdom.}

\author{Andrew~C.~Doherty}\email{doherty@physics.uq.edu.au}
\affiliation{School of Physical Sciences, The
  University of Queensland, Queensland 4072, Australia.}

\date{\today}

\begin{abstract}
In this paper we classify the four-qubit states that commute
with $U\ten{U}\ten{V}\ten{V}$, where $U$ and $V$ are arbitrary
members of the Pauli group. We characterize the set of
separable states for this class, in terms of a finite number of
entanglement witnesses. Equivalently, we characterize the
two-qubit, Bell-diagonal-preserving, completely positive maps
that are separable. These separable completely positive maps
correspond to protocols that can be implemented with stochastic
local operations assisted by classical communication (SLOCC).
This allows us to derive a complete set of SLOCC monotones for
Bell-diagonal states, which, in turn, provides the necessary
and sufficient conditions for converting one two-qubit state to
another by SLOCC.
\end{abstract}

\maketitle

\section{Introduction}
Entanglement has, unmistakeably, played a crucial role in many
quantum information processing tasks. Despite the various
separability criteria that have been developed, determining whether a
general multipartite mixed state is entangled is far from trivial. In
fact, computationally, the problem of deciding if a quantum state is
separable has been proven to be NP-hard~\cite{NP-hard}.

To date, separability of a general bipartite quantum state is
fully characterized only for dimension $2\times2$ and
$2\times3$~\cite{PPT}. For higher dimensional quantum systems,
there is no single criterion that is both necessary and
sufficient for separability. Nevertheless, for quantum states
that are invariant under some group of local unitary operators,
separability can often be determined relatively
easily~\cite{R.F.Werner:PRA:1989,M.Horodecki:PRA:1999,
K.G.H.Vollbrecht:PRA:2001,T.Eggeling:PRA:2001}.

On the other hand, it is often of interest in quantum
information processing to determine if a given state can be
transformed to some other desired state by local operations.
Indeed, convertibility between two (entangled) states using
local quantum operations assisted by classical communication
(LOCC) is closely related to the problem of quantifying the
entanglement associated to each quantum system. Intuitively,
one expects that a (single copy) entangled state can be locally
and deterministically transformed to a less entangled one but
not the other way round.

This intuition was made concrete in Nielsen's
work~\cite{M.A.Nielsen:PRL:1999} where he showed that a single
copy of a bipartite pure state $\ket{\Psi}$ can be locally and
deterministically transformed to another bipartite state
$\ket{\Phi}$, if and only if $\ket{\Phi}$ takes equal or lower
values for a set of functions known as entanglement
monotones~\cite{G.Vidal:JMP:2000}. One can, nevertheless, relax
the notion of convertibility by only requiring that the
conversion succeeds with some nonzero probability. Such
transformations are now known as stochastic LOCC
(SLOCC)~\cite{W.Dur:PRA:2000}. In this case, it was shown by
Vidal~\cite{G.Vidal:PRL:1999} that in the single copy scenario,
a pure state $\ket{\Psi}$ can be locally transformed to
$\ket{\Phi}$ with nonzero probability if and only if the
Schmidt rank of $\ket{\Psi}$ is higher than or equal to that of
$\ket{\Phi}$ (see also Ref.~\cite{W.Dur:PRA:2000}).

The analogous situation for mixed quantum states is not as well
understood even for two-qubit systems. If it were possible to
obtain a singlet state by SLOCC from a single copy of any mixed
state, it would be possible to convert any mixed state to any
other state~\cite{C.H.Bennett:PRA:1996}. However, as was shown
by Kent {\em et al.}~\cite{A.Kent:PRL:1999} (see also
Ref.~\cite{LX.Cen}), the best that one can do -- in terms of
increasing the entanglement of formation~\cite{S.Hill:PRL:1997}
-- is to obtain a Bell-diagonal state with higher but generally
non-maximal entanglement. In fact, apart from some rank
deficient states, this conversion process is known to be
invertible (with some
probability)~\cite{F.Verstraete:PRA:2001}. Hence, most
two-qubit states are known to be SLOCC equivalent to a
unique~\cite{fn:unique} Bell-diagonal state of
maximal~\cite{fn:maximal}
entanglement~\cite{A.Kent:PRL:1999,F.Verstraete:PRA:2001,F.Verstraete:PRA:2002}.

In this paper, we will complete the picture of two-qubit
convertibility under SLOCC by providing the necessary and
sufficient conditions for converting among Bell-diagonal
states. This characterization of the separable completely
positive maps (CPM) that take Bell diagonal states to Bell
diagonal states has other applications. Specifically, it was
required in the proof of our recent work~\cite{ANL} which
showed that all bipartite entangled states display a certain
kind of hidden non-locality~\cite{Hidden.Nonlocality}. (We show
that a bipartite quantum state violates the
Clauser-Horne-Shimony-Holt (CHSH) inequality~\cite{CHSH} after
local pre-processing with some non-CHSH violating ancilla state
if and only if the state is entangled.) Thus this paper
completes the proof of that result.

The structure of this paper is as follows. In
Sec.~\ref{Sec:SeparableStates}, we will start by characterizing
the set of separable states commuting with
$U\ten{U}\ten{V}\ten{V}$, where $U$ and $V$ are arbitrary
members of the Pauli group. Then, after reviewing the
one-to-one correspondence between separable maps and separable
quantum states in Sec.~\ref{Sec:SeparableMap},  we will derive,
in Sec.~\ref{Sec:BellMaps}, the full set of Bell-diagonal
preserving SLOCC transformations. A {\em complete} set of SLOCC
monotones are then derived in Sec.~\ref{Sec:Monotones} to
provide the necessary and sufficient conditions for converting
a Bell-diagonal state to another. This will then lead us to the
necessary and sufficient conditions that can be used to
determine if a two-qubit state can be converted to another
using SLOCC transformations. Finally, we conclude the paper
with a summary of results.

Throughout, the $(i,j)$-th entry of a matrix $W$ is denoted as
$[W]_{ij}$ (likewise $[\beta]_i$ for the $i$-th component of a
vector) whereas null entry in a matrix will be replaced by
$\cdot$ for ease of reading. Moreover, $\unit$ is the identity
matrix and $\Pi$ is used to denote a projector.

\section{Four-qubit Separable States with $U\ten{U}\ten{V}\ten{V}$
Symmetry}\label{Sec:SeparableStates}

Let us begin by reminding an important property of two-qubit
states which commute with all unitaries of the form $U\ten{U}$,
where $U$ are members of the Pauli group. The Pauli group is
generated by the Pauli matrices $\{\sigma_i\}_{i=x,y,z}$, and
has 16 elements. The representation $U\ten{U}$ decomposes onto
four one-dimensional irreducible representations, each acting
on the subspace spanned by one vector of the Bell basis
\begin{eqnarray}\label{Eq:BellBases}
  \ket{\Phi_{^1_2}} &\equiv& \frac{1}{\sqrt{2}}\left( \ket{00} \pm \ket{11} \right), \\
  \ket{\Phi_{^3_4}} &\equiv& \frac{1}{\sqrt{2}}\left( \ket{01} \pm \ket{10}
  \right).
\end{eqnarray}
This implies that~\cite{K.G.H.Vollbrecht:PRA:2001} any
two-qubit state which commutes with $U\ten{U}$ can be written
as $\rho=\sum_{i=1}^4 [r]_{i} \Pi_i$, where
$\Pi_i\equiv\proj{\Phi_i}$. With this information in mind, we
are now ready to discuss the case that is of our interest.

We would like to characterize the set of four-qubit states
which commute with all unitaries $U\ten{U}\ten{V}\ten{V}$,
where $U$ and $V$ are members of the Pauli group. Let us denote
this set of states by $\varrho$ and the state space of
$\rho\in\varrho$ as $\H\simeq
\H_{\A'}\ten\H_{\B'}\ten\H_{\A''}\ten\H_{\B''}$, where
$\H_{\A'}$, $\H_{\B'}$ etc. are Hilbert spaces of the
constituent qubits. In this notation, both the subsystems
associated with $\H_{\A'}\ten\H_{\B'}$ and that with
$\H_{\A''}\ten\H_{\B''}$ have $U\ten{U}$ symmetry and hence are
linear combinations of Bell-diagonal
projectors~\cite{K.G.H.Vollbrecht:PRA:2001}.

\begin{figure}[h!btp]
    \includegraphics[scale=1]{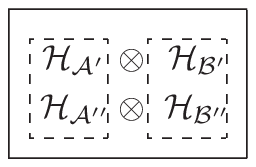}
    \caption{\label{Fig:StateSpace} A schematic diagram for the
    subsystems constituting $\rho$. Subsystems that are arranged in the
    same row in the diagram have $U\ten{U}$ symmetry and hence are
    represented by Bell-diagonal states~\cite{K.G.H.Vollbrecht:PRA:2001}
    (see text for details). In this paper, we are interested in states
    that are separable between subsystems enclosed in the two dashed boxes.}
\end{figure}

Our aim in this section is to provide a full characterization
of the set of $\rho$ that are separable between
$\H_\A\equiv\H_{\A'}\ten\H_{\A''}$ and
$\H_\B\equiv\H_{\B'}\ten\H_{\B''}$ (see
Fig.~\ref{Fig:StateSpace}). Throughout this section, a state is
said to be {\em separable} if and only if it is separable
between $\H_\A$ and $\H_\B$.

The symmetry of $\rho$ allows one to write it as a {\em
non-negative} combination of (tensored-) Bell projectors:
\begin{gather}\label{Eq:Rep}
    \rho=\sum_{i=1}^4\sum_{j=1}^4 [r]_{ij}\Pi_i\ten\Pi_j,
\end{gather}
where the Bell projector before and after the tensor product,
respectively, acts on $\H_{\A'}\ten\H_{\B'}$ and
$\H_{\A''}\ten\H_{\B''}$ (Fig.~\ref{Fig:StateSpace}). Thus, any
state $\rho\in\varrho$  can be represented in a compact manner,
via the corresponding $4\times4$ matrix $r$. More generally,
any operator $\mu$ acting on the same Hilbert space $\H$ and
having the same symmetry admits a $4\times4$ matrix
representation $M$ via:
\begin{gather}\label{Eq:Rep:General}
    \mu=\sum_{i=1}^4\sum_{j=1}^4 [M]_{ij}\Pi_i\ten\Pi_j,
\end{gather}
where $[M]_{ij}$ is now not necessarily non-negative. When
there is no risk of confusion, we will also refer to $r$ and
$M$, respectively, as a state and an operator having the
aforementioned symmetry.

Evidently, in this representation, an operator $\mu$ is
non-negative if and only if all entries in the corresponding
$4\times4$ matrix $M$ are non-negative. Notice also that by
appropriate local unitary transformation, one can swap any
$\Pi_i$ with any other $\Pi_j$, $j\neq i$ while keeping all the
other $\Pi_k$, $k\neq i,j$ unaffected. Here, the term {\em
local} is used with respect to the $\A$ and $\B$ partitioning.
Specifically, via the local unitary transformation
\begin{equation}
    V_{ij}\equiv\left\{ \begin{array}{c@{\quad:\quad}l}
    \half(\unit_2-\ii\sigma_z)\ten(\unit_2+\ii\sigma_z) & i=1,j=2,\\
    \half(\sigma_x+\sigma_z)\ten(\sigma_x+\sigma_z) & i=2,j=3,\\
    \half(\unit_2+\ii\sigma_z)\ten(\unit_2+\ii\sigma_z) & i=3,j=4,
    \end{array} \right.
\end{equation}
one can swap $\Pi_i$ and $\Pi_j$ while leaving all the other
Bell projectors unaffected. In terms of the corresponding
$4\times 4$ matrix representation, the effect of such local
unitaries on $\mu$ amounts to permutation of the rows and/or
columns of $M$. For brevity, in what follows, we will say that
two matrices $M$ and $M'$ are local unitarily equivalent if we
can obtain $M$ by simply permuting the rows and/or columns of
$M'$ and {\em vice versa}. A direct consequence of this
observation is that if $r$ represents a separable state, so is
any other $r'$ that is obtained from $r$ by independently
permuting any of its rows and/or columns.

Before we state the main result of this section, let us
introduce one more definition.
\begin{dfn}\label{dfn:rhos}
    Let $\P_s\subset\varrho$ be the convex hull of the states
    \begin{equation}\label{Eq:D0&G0}
        D_0\equiv \frac{1}{4}\left(
        \begin{array}{cccc}
        1 & \cdot & \cdot & \cdot \\
        \cdot & 1 & \cdot & \cdot \\
        \cdot & \cdot & 1 & \cdot \\
        \cdot & \cdot & \cdot & 1 \\
        \end{array}
        \right),\quad G_0\equiv \frac{1}{4}\left(
        \begin{array}{cccc}
        1 & 1 & \cdot & \cdot \\
        1 & 1 & \cdot & \cdot \\
        \cdot & \cdot & \cdot & \cdot \\
        \cdot & \cdot & \cdot & \cdot \\
        \end{array}
        \right),
    \end{equation}
    and the states that are local unitarily equivalent to these
    two.
\end{dfn}
Simple calculations show that with respect to the $\A$ and $\B$
partitioning, $D_0$, $G_0$ are separable~\cite{fn:separable}.
Hence, $\P_s$ is a separable subset of $\varrho$. The main
result of this section consists of showing the converse, and
hence the following theorem.
\begin{theorem}\label{Thm:SeparableUUVV}
    $\P_s$ is the set of states in $\varrho$ that are separable with
    respect to the $\A,\B$ partitioning.
\end{theorem}
Now, we note that $\P_s$ is a convex polytope. Its boundary is
therefore described by a finite number of
facets~\cite{B.Grunbaum:polytope}. Hence, to prove the above
theorem, it suffices to show that all these facets correspond to
valid entanglement witnesses. Denoting the set of facets by
$\mathcal{W}=\{W_i\}$. Then, using the software PORTA~\cite{PORTA},
the {\em nontrivial} facets were found to be
equivalent under local unitaries to one of the following:
\begin{widetext}
    \begin{gather}\label{Eq:W}
        W_1\equiv \left(
        \begin{array}{rrrr}
        1 & 1 & 1 &-1 \\
        1 & 1 & 1 &-1 \\
        1 & 1 & 1 &-1 \\
        -1 &-1 &-1 & 1 \\
        \end{array}
        \right),~W_2\equiv \left(
        \begin{array}{cccc}
        1 & 1 & \cdot & -1\\
        \cdot & \cdot & 1 & \cdot \\
        \cdot & \cdot & 1 & \cdot \\
        \cdot & \cdot & 1 & \cdot \\
        \end{array}
        \right),~W_3\equiv \left(
        \begin{array}{rrrr}
        3 & 3 & 1 &-1 \\
        3 &-1 & 1 & 3 \\
        1 & 1 & 3 & 1 \\
        -1 &-1 & 1 &-1 \\
        \end{array}
        \right),~W_4\equiv \left(
        \begin{array}{rrrr}
        3 & 3 & 1 &-1 \\
        3 &-1 & 1 & 3 \\
        3 &-1 & 1 &-1 \\
        1 & 1 &-1 & 1 \\
        \end{array}
        \right).
    \end{gather}
\end{widetext}
Apart from these, there is also a facet $W_0$ whose only
nonzero entry is $[W_0]_{11}=1$. $W_0$ and the operators local
unitarily equivalent to it give rise to  positive definite
matrices [c.f. Eq.~\eqref{Eq:ZW}], and thus correspond to
trivial entanglement witnesses. On the other hand, it is also
not difficult to verify that $W_1$ (and operators equivalent
under local unitaries) are decomposable and therefore demand
that $\rho_s$ remains positive semidefinite after partial
transposition. These are all the entanglement witnesses that
arise from the positive partial transposition (PPT)
requirement~\cite{PPT} for separable states.

To complete the proof of Theorem~\ref{Thm:SeparableUUVV}, it remains
to show that $W_2$, $W_3$, $W_4$ give rise to
Hermitian matrices
\begin{gather}\label{Eq:ZW}
    Z_{w,k}=\sum_{i=1}^4\sum_{j=1}^4
    [W_k]_{ij}~\left(\Pi_i\ten\Pi_j\right)
\end{gather}
that are valid entanglement witnesses, i.e.,
$\tr(\rho_sZ_{w,k})\ge 0$ for any separable $\rho_s\in\varrho$.
It turns out that this can be proved with the help of the
following lemma from Ref.~\cite{ACD:Extension}.

\begin{widetext}
\begin{lemma}\label{Lem:ProofOfWitnesses}
    For a given Hermitian matrix $Z_w$ acting on $\H_\A\ten\H_\B$, with
    $dim(\H_\A)=d_\A$ and $dim(\H_\B)=d_\B$, if there exists
    $m,n\in\mathbb{Z}^+$,  positive semidefinite $\Z$ acting on
    $\H_\A^{\ten m}\ten\H_\B^{\ten n}$ and a subset $s$ of the $m+n$
    tensor factors such that
    \begin{equation}\label{Eq:Z:Dfn}
        \pi_\A\ten\pi_\B~\left(\unit_{d_\A}^{\otimes  m-1}\ten
        Z_w\ten\unit_{d_\B}^{\ten n-1}\right)~\pi_\A\ten\pi_\B =
        \pi_\A\ten\pi_\B~\left(\Z^{{\rm T}_s}\right)~\pi_\A\ten\pi_\B,
    \end{equation}
    where $\pi_\A$ is the projector onto the symmetric subspace of
    $\H_\A^{\ten m}$ (likewise for $\pi_\B$) and $(.)^{{\rm T}_s}$
    refers to partial transposition with respect to the subsystem $s$,
    then $Z_w$ is a valid entanglement witness across $\HA$ and $\HB$,
    i.e., $\tr(Z_w\rho_\text{sep})\ge 0$ for any  state
    $\rho_\text{sep}$ that is separable with respect to the $\A$ and
    $\B$ partitioning.
\end{lemma}
\end{widetext}
\begin{proof}
Denote by $\A_k$ the subsystem associated with the $k$-th copy of
$\H_\A$ in $\H_\A^{\ten m}$; likewise for $\B_l$. To prove the above
lemma, let $\ket{\alpha}\in\H_\A$ and $\ket{\beta}\in\H_\B$ be
(unit) vectors, and for definiteness, let $s=\B_n$ then it follows
that
\begin{align*}
    &\bra{\alpha}\bra{\beta}~Z_w~\ket{\alpha}\ket{\beta}\\
    =&\bra{\alpha}^{\ten m}\bra{\beta}^{\ten n} \left(\unit_{d_\A}^{\ten
    m-1}\ten Z_w\ten\unit_{d_\B}^{\ten n-1}\right)\ket{\alpha}^{\ten
    m}\ket{\beta}^{\ten n}\\
    =&\bra{\alpha}^{\ten m}\bra{\beta}^{\ten n}\left[\pi_\A\ten
      \pi_\B\left(\Z^{{\rm T}_s}\right)\pi_\A\ten\pi_\B \right]\ket{\alpha}^{\ten
      m}\ket{\beta}^{\ten n}\\
    =&\bra{\alpha}^{\ten m}\bra{\beta}^{\ten n}\left(\Z^{{\rm
    T}_{\B_n}}\right)\ket{\alpha}^{\ten
      m}\ket{\beta}^{\ten n}\\
    =&\bra{\alpha}^{\ten m} \bra{\beta}^{\ten n-1}
    \ten\bra{\beta^*}~\Z~\ket{\alpha}^{\ten
      m}\ket{\beta}^{\ten n-1}\ten\ket{\beta^*}\\
    \ge &0,
\end{align*}
where $\ket{\beta^*}$ is the complex conjugate of
$\ket{\beta}$. We have made use of the identity
$\pi_\A\ket{\alpha}^{\ten m}=\ket{\alpha}^{\ten m}$ (likewise
for $\pi_\B$) in the second and third equality,
Eq.~\eqref{Eq:Z:Dfn} in the second equality, and the positive
semidefiniteness of $\Z$. To cater for general $s$, we just
have to modify the second to last line of the above computation
accordingly (i.e., to perform complex conjugation on all the
states in the set $s$) and the proof will proceed as before.
\end{proof}

More generally, let us remark that instead of having one $\Z$ on the
right hand side of Eq.~\eqref{Eq:Z:Dfn}, one can also have a sum of
different $\Z$'s, with each of them partial transposed with respect
to different subsystems $s$. Clearly, if the given $Z_w$ admits such
a decomposition, it is also an entanglement
witness~\cite{ACD:Extension}. For our purposes these
more complicated decompositions do not offer any advantage
over the simple decomposition given in Eq.~\eqref{Eq:Z:Dfn}.

By solving some appropriate semidefinite programs~\cite{SDP},
we have found that when $m=3$, $n=2$ and $s=\B_2$, there exist
some $\Z_k\ge0$, such that Eq.~\eqref{Eq:Z:Dfn} holds true for
each $k\in\{1,2,3,4\}$. Due to space limitations, the analytic
expression for these $\Z_k$'s will not be reproduced here but
are made available online at~\cite{url:z}. For $W_2$, the fact
that the corresponding $Z_{w,2}$ is a witness can even be
verified by considering $m=2$, $n=1$ and $s=\A_1$. In this
case, $d_\A=d_\B=4$. If we label the local basis vectors by
$\{\ket{i}\}_{i=0}^3$, the corresponding $\Z$ reads
\begin{gather*}
    \Z_2=\frac{1}{2}\sum_{i=1}^4\proj{z_i},\\
    \ket{z_1}=\ket{01,0}-\ket{02,3}+\ket{11,1}+\ket{13,3}+\ket{22,1}+\ket{23,0},\\
    \ket{z_2}=\ket{10,3}+\ket{11,2}+\ket{20,0}+\ket{22,2}-\ket{31,0}+\ket{32,3},\\
    \ket{z_3}=\ket{00,0}+\ket{02,2}+\ket{10,1}-\ket{13,2}+\ket{32,1}+\ket{33,0},\\
    \ket{z_4}=\ket{00,3}+\ket{01,2}-\ket{20,1}+\ket{23,2}+\ket{31,1}+\ket{33,3},
\end{gather*}
where we have separated $\A$'s degree of freedom from $\B$'s
ones by comma~\cite{fn:Reorder}. This completes the proof for
Theorem~\ref{Thm:SeparableUUVV}.

\section{SLOCC Convertibility of Bell-Diagonal
States}\label{Sec:SLOCC-BDS}

An immediate corollary of the characterization given in
Sec.~\ref{Sec:SeparableStates} is that we now know exactly the
set of Bell-diagonal preserving transformations that can be
performed locally on a Bell-diagonal state. In this section, we
will make use of the Choi-Jamio\l{k}owski
isomorphism~\cite{Jamiolkowski}, i.e., the one-to-one
correspondence between completely positive map (CPM) and
quantum state, to make these SLOCC transformations explicit.
This will allow us to derive a complete set of SLOCC
monotones~\cite{G.Vidal:JMP:2000} which, in turn, serve as a
set of necessary and sufficient conditions for converting one
Bell-diagonal state to another.

\subsection{Separable Maps and SLOCC}\label{Sec:SeparableMap}

Now, let us recall some well-established facts about CPM. To
begin with, a separable CPM, denoted by $\E_s$ takes the
following form~\cite{E.M.Rains:9707002,V.Vedral:PRA:1998}
\begin{equation}\label{Eq:SeparableMap}
    \E_s:\rho\to\sum_{i=1}^n(A_i\ten B_i)~\rho~(A_i^\dag\ten B_i^\dag),
\end{equation}
where $\rho$ acts on $\H_{\Ai}\ten\H_{\Bi}$, $A_i$ acts on
$\H_{\Ai}$, $B_i$ acts on $\H_{\Bi}$~\cite{fn:Kraus}. If,
moreover,
\begin{equation}\label{Eq:TracePreserving}
    \sum_i \left(A_i\ten{B_i}\right)^\dag
    \left(A_i\ten{B_i}\right)=\unit,
\end{equation}
the map is trace-preserving, i.e., if $\rho$ is normalized, so
is the output of the map $\E_s(\rho)$. Equivalently, the
trace-preserving condition demands that the transformation from
$\rho$ to $\E_s(\rho)$ can always be achieved with certainty.
It is well-known that all LOCC transformations are of the form
Eq.~\eqref{Eq:SeparableMap} but the converse is not
true~\cite{C.H.Bennett:PRA:1999}.

However, if we allow the map $\rho\to\E_s(\rho)$ to fail with
some probability $p<1$, the transformation from $\rho$ to
$\E_s(\rho)$ can always be implemented probabilistically via
LOCC. In other words, if we do not impose
Eq.~\eqref{Eq:TracePreserving}, then
Eq.~\eqref{Eq:SeparableMap} represents, up to some
normalization constant, the most general LOCC possible on a
bipartite quantum system. These are the SLOCC
transformations~\cite{W.Dur:PRA:2000}.

To make a connection between the set of SLOCC transformations
and the set of states that we have characterized in
Sec.~\ref{Sec:SeparableStates}, let us also recall the
Choi-Jamio{\l}kowski isomorphism~\cite{Jamiolkowski} between
CPM and quantum states: for every (not necessarily separable)
CPM $\E:\H_{\Ai}\ten\H_{\Bi}\to\H_{\Ao}\ten\H_{\Bo}$ there is a
unique -- again, up to some positive constant $\alpha$ --
quantum state $\rho_\E$ corresponding to $\E$:
\begin{equation}\label{Eq:JamiolkowskiState}
    \rho_\E=\alpha~\E\otimes \mathcal{I} \left(\ket{\Phi^+}_{\Ai}\bra{\Phi^+}\ten
    \ket{\Phi^+}_{\Bi}\bra{\Phi^+}\right),
\end{equation}
where
$\ket{\Phi^+}_{\Ai}\equiv\sum_{i=1}^{d_{\Ai}}\ket{i}\ten\ket{i}$
is the unnormalized maximally entangled state of dimension
$d_{\Ai}$ (likewise for $\ket{\Phi^+}_{\Bi}$). In
Eq.~\eqref{Eq:JamiolkowskiState}, it is understood that $\E$
only acts on half of $\ket{\Phi^+}_{\Ai}$ and half of
$\ket{\Phi^+}_{\Bi}$. Clearly, the state $\rho_\E$ acts on a
Hilbert space of dimension $d_{\Ai}\times d_{\Ao}\times
d_{\Bi}\times d_{\Bo}$, where $d_{\Ao}\times d_{\Bo}$ is the
dimension of $\H_{\Ao}\ten\H_{\Bo}$.

Conversely, given a state $\rho_\E$ acting on
$\H_{\Ao}\ten\H_{\Bo}\ten\H_{\Ai}\ten\H_{\Bi}$, the
corresponding action of the CPM $\E$ on some $\rho$ acting on
$\H_{\Ai}\ten\H_{\Bi}$ reads:
\begin{equation}\label{Eq:State->CPM}
    \E(\rho)=\frac{1}{\alpha}\tr_{\Ai\Bi}\left[\rho_\E\left(\unit_{\Ao\Bo}\ten\rho\t\right)\right],
\end{equation}
where $\rho\t$ denote transposition of $\rho$ in some local
bases of $\H_{\Ai}\ten\H_{\Bi}$. For a trace-preserving CPM, it
then follows that we must have
$\tr_{\Ao\Bo}(\rho_\E)=\alpha\unit_{\Ai\Bi}$. A point that
should be emphasized now is that $\E$ is a separable map [c.f.
Eq.~\eqref{Eq:SeparableMap}] if and only if the corresponding
$\rho_\E$ given by Eq.~\eqref{Eq:JamiolkowskiState} is
separable across $\H_{\Ai}\ten\H_{\Ao}$ and
$\H_{\Bi}\ten\H_{\Bo}$~\cite{J.I.Cirac:PRL:2001}. Moreover, at
the risk of repeating ourselves, the map $\rho\to\E(\rho)$
derived from a separable $\rho_\E$ can always be implemented
locally, although it may only succeed with some (nonzero)
probability. Hence, if we are only interested in
transformations that can be performed locally, and not the
probability of success in mapping $\rho\to\E(\rho)$, the
normalization constant $\alpha$ as well as the normalization of
$\rho_\E$ becomes irrelevant. This is the convention that we
will adopt for the rest of this section.

\subsection{Bell-diagonal Preserving SLOCC Transformations}
\label{Sec:BellMaps}

We shall now apply the isomorphism to the class of states
$\varrho$ that we have characterized in
Sec.~\ref{Sec:SeparableStates}. In particular, if we identify
$\Ai$, $\Ao$, $\Bi$ and $\Bo$ with, respectively, $\A''$,
$\A'$, $\B''$ and $\B'$, it  follows from Eq.~\eqref{Eq:Rep}
and Eq.~\eqref{Eq:State->CPM} that for any two-qubit state
$\rho_{\rm in}$, the action of the CPM derived from
$\rho\in\varrho$ reads:
\begin{equation}
    \E:\rho_{\rm in}\to\rho_{\rm out}\propto \sum_{i,j}[r]_{ij}
    \tr\left(\rho\t_{\rm in}\Pi_j\right)\Pi_i.
\end{equation}
Hence, under the action of $\E$, any $\rho_{\rm in}$ is
transformed to another two-qubit state that is diagonal in the
Bell basis, i.e., a Bell-diagonal state. In particular, for a
Bell-diagonal $\rho_{\rm in}$, i.e.,
\begin{gather}
    \rho_{\rm in}=\sum_k[\beta]_k\Pi_k,\nonumber\\
    [\beta]_k\ge0,\quad \sum_k[\beta]_k=1,
\end{gather}
the map outputs another Bell-diagonal state
\begin{equation}
    \rho_{\rm out}=\E(\rho_{\rm in})\propto\sum_{i,j}
    [\beta]_j[r]_{ij}\Pi_i.
\end{equation}
It is worth noting that for general $\rho_\E\in\varrho$,
$\tr_{\A'\B'}\rho_\E$ is not proportional to the identity
matrix, therefore some of the CPMs derived from
$\rho\in\varrho$ are intrinsically
non-trace-preserving~\cite{fn:Stochasticity}.

By considering the convex cone~\cite{fn:cone} of separable
states $\P_s$ that we have characterized in
Sec.~\ref{Sec:SeparableStates}, we therefore obtain the entire
set of Bell-diagonal preserving SLOCC transformations. Among
them, we note that the extremal maps, i.e., those derived from
Eq.~\eqref{Eq:D0&G0}, admit simple physical interpretations and
implementations. In particular, the extremal separable map for
$D_0$, and the maps that are related to it by local unitaries,
correspond to permutation of the input Bell projectors $\Pi_i$
-- which can be implemented by performing appropriate local
unitary transformations. The other kind of extremal separable
map, derived from $G_0$, corresponds to making a measurement
that determines if the initial state is in a subspace spanned
by a given pair of Bell states and if successful discarding the
input state and replacing it by an equal but incoherent mixture
of two of the Bell states. This operation can be implemented
locally since the equally weighted mixture of two Bell states
is a separable state and hence  both the measurement step and
the state preparation step can be implemented locally.

\begin{figure}[h!btp]
    \includegraphics[scale=0.6]{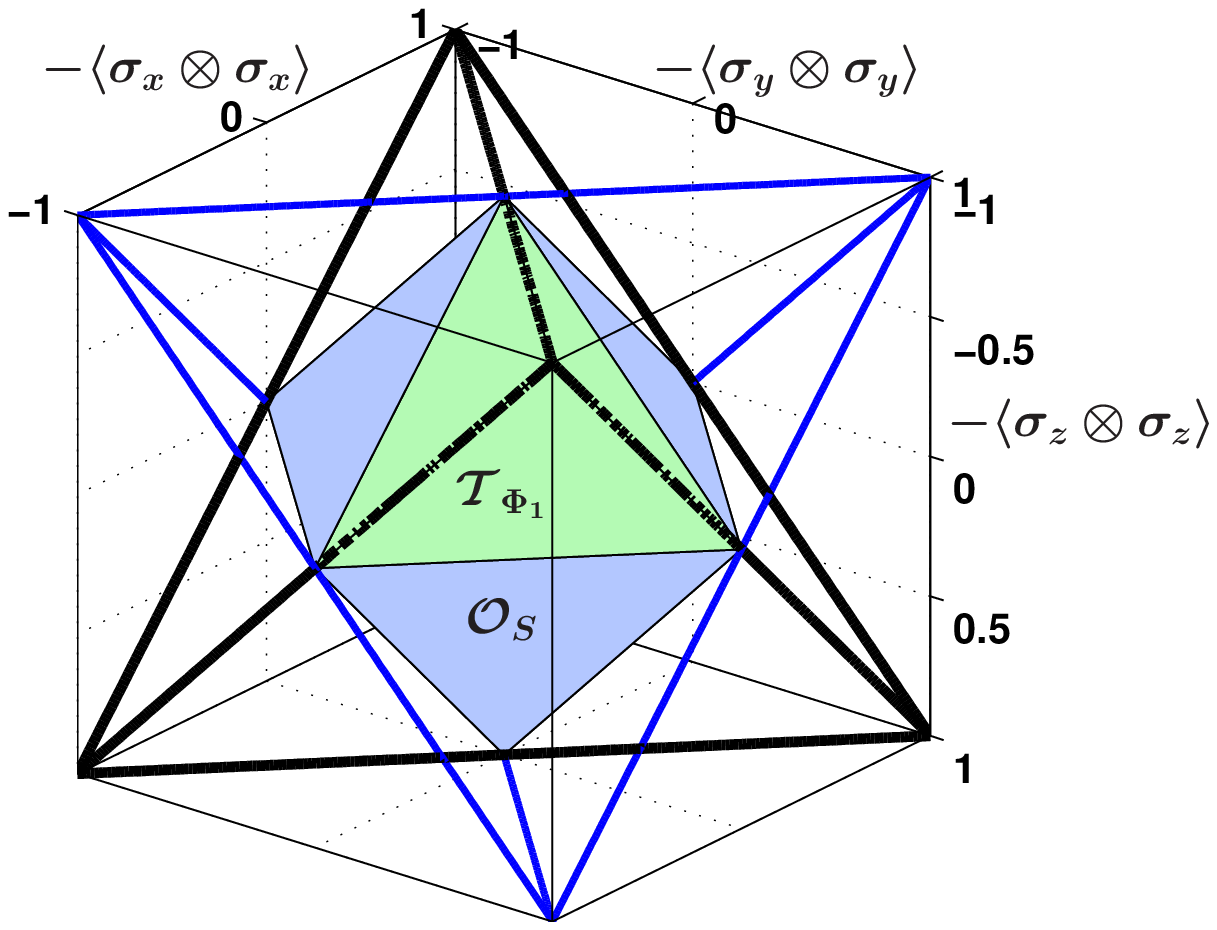}
    \caption{\label{Fig:BDS}
    (Color online) State space for Bell-diagonal states. The set of
    physical states is the tetrahedron $\T_0$ whose edges are marked with
    thick (black) lines whereas the set having positive partial transpose
    is another tetrahedron whose edges are marked with thinner (blue)
    lines. The intersection of the two tetrahedra gives rise to the
    octahedron $\O_S$ (blue) which is the set of separable Bell-diagonal
    states. Entangled Bell-diagonal states satisfying
    Eq.~\eqref{Eq:LambdaOrder} are contained within the tetrahedron
    $\T_{\Phi_1}$ (green), which is discussed further in
    Fig.~\ref{Fig:Plambda}.
    }
\end{figure}

\subsection{Complete Set of SLOCC Monotones for Bell-diagonal States}
\label{Sec:Monotones}

Now, let us make use of the above characterization to derive a
{\em complete} set of {\em non-increasing} SLOCC monotones for
Bell-diagonal states. To begin with, we recall that the set of
normalized Bell-diagonal states forms a tetrahedron $\T_0$ in
$\mathbb{R}^3$, and the set of separable Bell-diagonal states
forms an octahedron $\mathcal{O}_S$ (see Fig.~\ref{Fig:BDS})
that is contained in $\T_0$~\cite{K.G.H.Vollbrecht:PRA:2001}.
We will follow Ref.~\cite{K.G.H.Vollbrecht:PRA:2001} and use
the expectation values
$(-\langle\sigma_x\ten\sigma_x\rangle,-\langle\sigma_y\ten\sigma_y\rangle,
-\langle\sigma_z\ten\sigma_z\rangle)$ as the coordinates of
this three-dimensional space. The coordinates of the four Bell
states $\{\ket{\Phi_i}\}_{i=1}^4$ are then $(-1,1,-1)$,
$(1,-1,-1)$,$(-1,-1,1)$ and $(1,1,1)$ respectively.

Since Bell-diagonal states are convex mixtures of the four Bell
projectors, we may also label any point in the state space of
Bell-diagonal states by a four-component weight vector
$\vec{\lambda}=(\lambda_1,\lambda_2,\lambda_2,\lambda_4)$ such
that the corresponding Bell-diagonal state reads
\begin{equation}\label{Eq:BDS}
    \rbd(\vl)=\sum_i^4\lambda_i\Pi_i.
\end{equation}
Moreover, as remarked above, we can apply local unitary
transformation to swap any of the two Bell projectors while
leaving others unaffected. Thus, without loss of generality, we
will restrict our attention to Bell-diagonal states such that
\begin{equation}\label{Eq:LambdaOrder}
    \lambda_1\ge\lambda_2\ge\lambda_3\ge\lambda_4,
\end{equation}
and determine when it is possible to transform between two such
states under SLOCC.

\begin{figure}[h!btp]
    \includegraphics[scale=0.6]{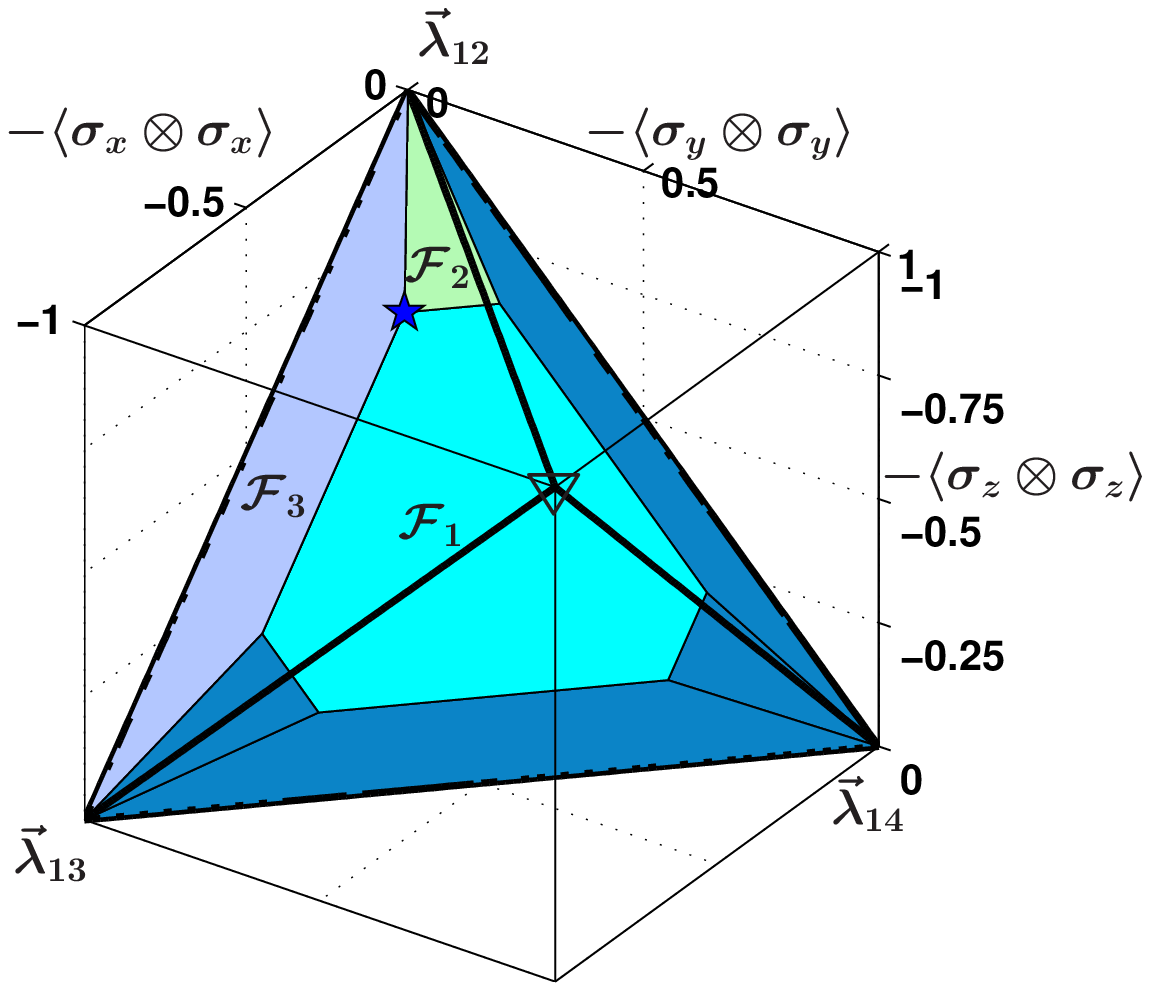}
    \caption{\label{Fig:Plambda}
    (Color online) Tetrahedron $\T_{\Phi_1}$ (with edges marked with
    thick black lines) is the set of Bell-diagonal states with
    $\ld_1\ge1/2$. Its four vertices are the Bell state $\ket{\Phi_1}$
    (marked with a $\bigtriangledown$) and the three separable states
    $\vl_{12}$, $\vl_{13}$ and $\vl_{14}$. Within $\T_{\Phi_1}$ is the
    convex polytope $\P_\ld$, which are points within $\T_{\Phi_1}$ that
    can be obtained from $\vl$ (marked with a $\star$) by performing
    SLOCC transformations. Three of the facets of $\P_\ld$, namely
    $\F_1$, $\F_2$ and $\F_3$ are shown with cyan, green and light purple
    colors, respectively; the other facets of $\P_\ld$ are shown with
    blue color.
    }
\end{figure}

Clearly, any (entangled) Bell-diagonal state can be transformed
to any separable Bell-diagonal state via SLOCC -- one can
simply discard the original Bell-diagonal state and prepare the
separable state using LOCC. Also separable Bell-diagonal states
can only be transformed among themselves with SLOCC.

What about transformations among entangled Bell-diagonal
states? To answer this question, we shall adopt the following
strategy. Firstly, we will clarify -- in relation to
Fig.~\ref{Fig:BDS} -- the set of entangled Bell-diagonal states
satisfying Eq.~\eqref{Eq:LambdaOrder}. Then, we will make use
of the characterization obtained in Sec.~\ref{Sec:BellMaps} to
determine the set of states that can be obtained from SLOCC
transformations when we have an input (entangled) state
satisfying Eq.~\eqref{Eq:LambdaOrder}. After that, we will
restrict our attention to the subset of these output states
satisfying Eq.~\eqref{Eq:LambdaOrder}. Once we have got this, a
simple set of necessary and sufficient conditions can be
derived to determine if an entangled Bell-diagonal state can be
converted to another.

We now take a closer look at the set of entangled Bell-diagonal
states, in particular those that satisfy
Eq.~\eqref{Eq:LambdaOrder}. In Fig.~\ref{Fig:BDS}, the set of
entangled Bell-diagonal states is the relative complement of
the (blue) octahedron $\O_S$ in the tetrahedron $\T_0$. In this
set, those points that satisfy Eq.~\eqref{Eq:LambdaOrder} are a
strict {\em subset} contained in the (green) tetrahedron
$\T_{\Phi_1}$, which has the Bell state $\ket{\Phi_1}$ and the
three mixed separable states
\begin{equation}\label{Eq:BDS:Separable:1+i}
    \rho_{1i}=\half\left(\proj{\Phi_1}+\proj{\Phi_i}\right),
    \quad i=2,3,4,
\end{equation}
as its four vertices. In terms of weight vectors, the three
separable vertices read
\begin{equation}\label{Eq:Tphi}
    \vl_{12}=\half\left(\begin{array}{c}
    1\\1\\ \cdot \\ \cdot\end{array}\right),
    \vl_{13}=\half\left(\begin{array}{c}
    1\\\cdot\\1\\\cdot\end{array}\right),
    \vl_{14}=\half\left(\begin{array}{c}
    1\\\cdot\\\cdot\\1\end{array}\right).
\end{equation}
$\T_{\Phi_1}$ is the set of Bell-diagonal states satisfying
$\ld_1\ge1/2$ which includes both entangled states (denoted by
$\TE$) and separable states (denoted by $\F_0$). For the
purpose of subsequent discussion, it is important to note that
every entangled state satisfying Eq.~\eqref{Eq:LambdaOrder} is
in $\TE$ but not every state in $\TE$ satisfies
Eq.~\eqref{Eq:LambdaOrder}.

Now, let us consider an entangled Bell-diagonal state
$\rbd(\vl)$ with weight vector
\begin{equation}\label{Eq:Dfn:Lambda}
    \vl=\left(\begin{array}{c}\ld_1\\\ld_2\\\ld_3\\\ld_4\end{array}\right)
\end{equation}
satisfying Eq.~\eqref{Eq:LambdaOrder}. Note from the above
discussion that $\vl\in\TE$. Recall that our goal is to
determine the set of (entangled) output states -- satisfying
Eq.~\eqref{Eq:LambdaOrder} -- which can be obtained from $\vl$
via SLOCC. To achieve that, we will begin by first determining
the set of output weight vectors $\{\vl'\}$ which are in the
superset $\T_{\Phi_1}$.

In particular, we note that under extremal SLOCC
transformations associated with $G_0$, and the operators local
unitarily equivalent to it [c.f. Eq.~\eqref{Eq:D0&G0} and
Sec.~\ref{Sec:BellMaps}], $\vl$ can be brought into any of the
separable states $\{\rho_{1i}\}_{i=2}^4$ [c.f.
Eqs.~\eqref{Eq:BDS:Separable:1+i} and \eqref{Eq:Tphi}].
Similarly, under extremal SLOCC transformations associated with
$D_0$, and the operators local unitarily equivalent to it,
$\vl$ can be brought into any of the following entangled
Bell-diagonal states by permuting the weights associated with
some of the Bell projectors:
\begin{gather}
    \vl_{(34)}=\left(\begin{array}{c}\ld_1\\\ld_2\\\ld_4\\\ld_3\end{array}\right),
    \vl_{(324)}=\left(\begin{array}{c}\ld_1\\\ld_3\\\ld_4\\\ld_2\end{array}\right),
    \vl_{(24)}=\left(\begin{array}{c}\ld_1\\\ld_4\\\ld_3\\\ld_2\end{array}\right),\nonumber\\
    \vl_{(234)}=\left(\begin{array}{c}\ld_1\\\ld_4\\\ld_2\\\ld_3\end{array}\right),
    \label{Eq:ExtremePoints}\vl_{(23)}=\left(\begin{array}{c}\ld_1\\\ld_3\\\ld_2\\\ld_4\end{array}\right).
\end{gather}
Evidently, any convex combinations of the vectors listed in
Eq.~\eqref{Eq:Tphi}, Eq.~\eqref{Eq:Dfn:Lambda} and
Eq.~\eqref{Eq:ExtremePoints} are also attainable from $\vl$
using (non-extremal) SLOCC. Moreover, within $\T_{\Phi_1}$,
only convex combinations of these states, denoted by $\P_\ld$,
are attainable from $\vl$ using SLOCC. $\P_\ld$  is thus a
convex polytope with vertices given by the union of vectors
listed in Eq.~\eqref{Eq:Tphi}, Eq.~\eqref{Eq:Dfn:Lambda} and
Eq.~\eqref{Eq:ExtremePoints}.

Then, to determine if $\vl$ can be transformed to another
$\vlp\in\T_{\Phi_1}$ amounts to deciding if $\vlp\in\P_\ld$. It is a
well known fact that a convex polytope can also be described by a
finite set of inequalities that are associated with each of the
facets of the polytope~\cite{B.Grunbaum:polytope}. Therefore, the
above task can be done, for example, by checking if $\vlp$ satisfies
all the linear equalities defining the polytope $\P_\ld$.

Our real interest, however, is in the set of entangled Bell-diagonal
states satisfying Eq.~\eqref{Eq:LambdaOrder}. With some thought, it
should be clear that this simplifies the problem at hand so that we
will only need to check that $\vlp$ satisfies all the inequalities
(facets) that contain $\vl$. From Fig.~\ref{Fig:Plambda},
it can be seen that only three facets of $\P_\ld$ contain
$\vl$. These are $\F_1=\text{conv}\{\vl, \vl_{(34)}, \vl_{(324)},
\vl_{(24)}, \vl_{(234)}, \vl_{(23)}\}$,
$\F_2=\text{conv}\{\vl,\vl_{12},\vl_{(34)}\}$ and
$\F_3=\text{conv}\{\vl,\vl_{12},\vl_{13},\vl_{(23)}\}$, where
$\text{conv}\{.\}$ represents the convex hull formed by the set of
points in $\{.\}$~\cite{B.Grunbaum:polytope}.

Recall that each vector $\vl_{(.)}$ listed in
Eq.~\eqref{Eq:ExtremePoints} is obtained by performing the
appropriate permutation $(.)$ on all but the first component of
$\vl$. Hence $\F_1$ is a facet of constant $\ld_1$. After some
simple algebra, the inequalities associated with
$\F_2$~\cite{fn:F2} and $\F_3$~\cite{fn:F3} can be shown to be,
respectively,
\begin{gather}\label{Eq:F2}
    \F_2:\frac{\ld_3+\ld_4}{\ld_1-\ld_2}
    \left(\langle\sigma_x\ten\sigma_x\rangle-\langle\sigma_y\ten\sigma_y\rangle\right)
    +\langle\sigma_z\ten\sigma_z\rangle\le1,\\
    \label{Eq:F3}\F_3:\langle\sigma_x\ten\sigma_x\rangle+\langle\sigma_z\ten\sigma_z\rangle
    -\frac{1-2\ld_1+2\ld_4}{1-2\ld_2-2\ld_3}\langle\sigma_y\ten\sigma_y\rangle\le1.
\end{gather}
Imposing the requirement that $\vlp$ satisfies these inequalities
gives, respectively,
\begin{gather*}
    \frac{1-2\ld_2}{\ld_3+\ld_4} \ge \frac{1-2\ldp_2}{\ldp_3+\ldp_4},
\end{gather*}
and
\begin{gather*}
    \frac{1-2\ld_2-2\ld_3}{\ld_4}\ge
    \frac{1-2\ldp_2-2\ldp_3}{\ldp_4}.
\end{gather*}
Together with the requirement imposed by $\F_1$, we see that by
defining
\begin{align}
    E_1(\vl)&\equiv\ld_1, \\
    E_2(\vl)&\equiv\frac{1-2\ld_2}{\ld_3+\ld_4},\\
    E_3(\vl)&\equiv\frac{1-2\ld_2-2\ld_3}{\ld_4},
\end{align}
the intercovertibility between two entangled Bell-diagonal states
can be succinctly summarized in the following theorem.
\begin{theorem}\label{Thm:Monotones}
    Let $\rho$ and $\rho'$ be two entangled Bell-diagonal states with,
    respectively, weight vectors $\vl$ and $\vl'$ satisfying
    Eq.~\eqref{Eq:LambdaOrder}. Transformation from $\rho$ to $\rho'$
    via SLOCC is possible iff
    \begin{align}
        E_1(\vl)\ge E_1(\vlp),\label{Eq:E1}\\
        E_2(\vl)\ge E_2(\vlp),\label{Eq:E2}\\
        E_3(\vl)\ge E_3(\vlp)\label{Eq:E3}.
    \end{align}
    In other words, $\{E_i(\vl)\}_{i=1}^3$ is a complete set of SLOCC
    monotones for entangled Bell-diagonal states satisfying
    Eq.~\eqref{Eq:LambdaOrder}.
\end{theorem}

\section{SLOCC Convertibility of Two-Qubit States}

With Theorem~\ref{Thm:Monotones}, it is just another small step
to determine if a two-qubit state $\rho$ can be converted to
another, say $\rho'$ using SLOCC. To this end, let us first
recall the following definition from
Ref.~\cite{W.Dur:PRA:2000}.
\begin{dfn}
    Two states $\rho$ and $\rho'$ are said to be SLOCC
    equivalent if $\rho$ can be converted to $\rho'$ via SLOCC
    with nonzero probability and vice versa.
\end{dfn}
Next, we recall the following theorem, which can be deduced
from Theorem 1 in Ref.~\cite{F.Verstraete:PRA:2002} (see also
Theorems 1--3 in Ref.~\cite{F.Verstraete:PRA:2001}).
\begin{theorem}
    A two-qubit state $\rho$ is SLOCC equivalent to either (1) a unique
    Bell-diagonal state satisfying Eq.~\eqref{Eq:LambdaOrder}, (2) a
    separable state, or (3) a (normalized) non-Bell-diagonal state of
    the form:
    \begin{equation}
    \label{Eq:quasi}
        \rho_{\rm ND}=\frac{1}{4}\left(\begin{array}{cccc}
        2 & \cdot & \cdot & \cdot \\
        \cdot & 1 & 2b & \cdot \\
        \cdot & 2b & 1 & \cdot \\
        \cdot & \cdot & \cdot & \cdot
        \end{array}\right),
    \end{equation}
    where $\rho_{\rm ND}$ is expressed in the standard product
    basis and $b\le\half$ is unique.
\end{theorem}

Moreover, as was shown in Ref.~\cite{F.Verstraete:PRA:2001}, the
unique Bell-diagonal state in case (1) is the state with maximal
entanglement that can be obtained from the original two-qubit state
using SLOCC. The two-qubit state associated with case (2) is clearly
a separable one since a separable state is, and can only be, SLOCC
equivalent to another separable state.

The situation for case (3) is somewhat more complicated and the
original two-qubit states associated with this case are either of
rank 3 or 2 (in the case of $b=1/2$)
~\cite{LX.Cen,F.Verstraete:PRA:2001,F.Verstraete:PRA:2002}. By very
inefficient SLOCC transformations --
quasi-distillation~\cite{MPR.Horodecki:PRA:1999} -- the entanglement
in the equivalent state $\rho_\text{ND}$ can be maximized by
converting it into the following Bell-diagonal state:
\begin{equation}
    \rho_{\rm ND}'=\half\left(\begin{array}{cccc}
    \cdot & \cdot & \cdot & \cdot \\
    \cdot & 1 & -2b & \cdot \\
    \cdot & -2b & 1 & \cdot \\
    \cdot & \cdot & \cdot & \cdot
    \end{array}\right).
\end{equation}
However, it remains unclear from existing
results~\cite{LX.Cen,MPR.Horodecki:PRA:1999,
F.Verstraete:PRA:2001,F.Verstraete:PRA:2002} if this process is
reversible~\cite{fn:reverse}. In this regard, we have found that the
reverse process can indeed be carried out via a separable map with
two terms involved in the Kraus decomposition. In particular, a
possible form of the Kraus operators associated with this separable
map reads [Eq.~\eqref{Eq:SeparableMap}]:
\begin{gather*}
    A_1=\left(\begin{array}{cc} -2b+\sqrt{1+4b^2} & -1/2\\
    1 & \cdot \end{array}\right),\quad
    B_1=\left(\begin{array}{cc} 1 & 1/2\\
    1 & \cdot \end{array}\right),\\
    A_2=\left(\begin{array}{cc} 2b-\sqrt{1+4b^2} & 1/2\\
    1 & \cdot \end{array}\right),\quad
    B_2=\left(\begin{array}{cc} 1 & 1/2\\
    -1 & \cdot \end{array}\right).
\end{gather*}
Thus, a two-qubit state that is SLOCC equivalent to $\rho_\text{ND}$
is also SLOCC equivalent to a unique Bell-diagonal state
$\rho_\text{ND}'$. By further local unitary transformation, we can
bring $\rho_\text{ND}'$ into a form that satisfies
Eq.~\eqref{Eq:LambdaOrder}. Hence, this leads us to the following
theorem.

\begin{theorem}
    All entangled two-qubit states are SLOCC equivalent to a unique
    Bell-diagonal state satisfying Eq.~\eqref{Eq:LambdaOrder}.
\end{theorem}

With this theorem, one can now readily determine if an {\em
entangled} two-qubit state $\rho$ can be converted to another,
say, $\rho'$, using SLOCC. For that matter, let $\rbd(\vl)$ and
$\rbd(\vl')$ be, respectively, the unique Bell-diagonal state
satisfying Eq.~\eqref{Eq:LambdaOrder} that is SLOCC equivalent
to $\rho$ and $\rho'$. Then, it follows from
Theorem~\ref{Thm:Monotones} that $\rho$ can be transformed to
$\rho'$ using SLOCC if and only if the corresponding weight
vectors of the associated Bell-diagonal states $\vl$ and $\vl'$
satisfy Eqs.~\eqref{Eq:E1}-\eqref{Eq:E3}. In other words, the
SLOCC convertibility of two two-qubit states can be decided via
the following theorem.

\begin{theorem}
    Let $\rbd(\vl)$ and $\rbd(\vl')$ be, respectively, the Bell-diagonal
    state satisfying Eq.~\eqref{Eq:LambdaOrder} that is SLOCC equivalent
    to $\rho$ and $\rho'$. $\rho$ can be locally transformed onto $\rho'$
    with nonzero probability if and only if (1) $\rho'$ is separable or
    (2) $\rho$ is entangled and the associated weight vectors $\vl$ and
    $\vl'$ satisfy Eqs.~\eqref{Eq:E1}-\eqref{Eq:E3}.
\end{theorem}

Schematically, if neither $\rho$ nor $\rho'$ are separable and if
Eqs.~\eqref{Eq:E1}-\eqref{Eq:E3} are satisfied, then one possible
way of transforming $\rho$ to $\rho'$ via SLOCC is by performing the
following chain of conversions:
\begin{equation*}
    \rho\to\rbd(\vl)\to\rbd(\vl')\to\rho',
\end{equation*}
whereas if any one of Eqs.~\eqref{Eq:E1}-\eqref{Eq:E3} is not
satisfied, then
\begin{equation*}
    \rho\not\to\rho'.
\end{equation*}

\vspace{1cm}
\section{Discussion and Conclusion}

In this paper, we have investigated the bi-separability of the set
of four-qubit states commuting with $U\ten{U}\ten{V}\ten{V}$ where
$U$ and $V$ are arbitrary members of the Pauli group. These are
essentially convex combination of two (not necessarily identical)
copies of Bell states. Evidently, these states are all separable
across the two copies. For the other bi-partitioning, we have found
that the separable subset is a convex polytope and hence can be
described by a finite set of entanglement witnesses.

Equivalently, this characterization has also given us the complete
set of separable, Bell-diagonal preserving, completely positive
maps. This has enabled us to derive a complete set of SLOCC
monotones for Bell-diagonal states, which can be used to determine
if a Bell-diagonal state can be converted to another using SLOCC.

We have then supplemented the result on SLOCC equivalence presented
in Refs.~\cite{F.Verstraete:PRA:2001,F.Verstraete:PRA:2002} to
arrive at the conclusion that all entangled two-qubit states are
SLOCC equivalent to a unique Bell-diagonal state. Combining this
with the SLOCC monotones that we have derived immediately leads us
to some simple necessary and sufficient criteria on the SLOCC
convertibility between two-qubit states.

\begin{acknowledgements}
We would like to thank Guifr\'e Vidal and Frank Verstraete for
helpful discussions. This work is supported by the EU Project
QAP (IST-3-015848) and the Australian Research Council.
\end{acknowledgements}


\begin{thebibliography}{99}
\bibitem{fn:maximal} Maximal, in the sense that no SLOCC
    transformations can bring the state to another one with
    higher entanglement of formation.
\bibitem{fn:separable} Their separability can be veirfied, for
    example, by writing these operators in full in the product
    basis via Eq.~\eqref{Eq:Rep:General} and showing that they
    admit convex decomposition in terms of separable states.
    Alternatively, via the Choi-Jamio{\l}kowski isomorphism
    that will be discussed later in Sec.~\ref{Sec:SeparableMap}
    and the remarks made towards the end of
    Sec.~\ref{Sec:BellMaps}, one can also see that these
    matrices correspond to separable states.
\bibitem{fn:Reorder} Note that to verify $\Z_2$ against
    Eq.~\eqref{Eq:Z:Dfn}, one should also rewrite $Z_{w,2}$
    obtained in Eq.~\eqref{Eq:ZW} in the appropriate
    tensor-product basis such that $Z_{w,2}$ acts on
    $\H_{\A'}\ten\H_{\A''}\ten\H_{\B'}\ten\H_{\B''}$.
\bibitem{fn:Kraus} Following Kraus' work on CPM~\cite{Kraus},
    this specific form of the CPM is also known as a Kraus
    decomposition of the CPM, with each $A_i\ten B_i$ in the
    sum conventionally called the Kraus operator associated
    with the CPM.
\bibitem{fn:Stochasticity} The $\rho_\E$ derived from $G_0$ in
    Eq.~\eqref{Eq:D0&G0} is an example of this sort. In fact,
    in this case, if the input state has no support on $\Pi_1$
    nor $\Pi_2$, the map always outputs the zero matrix.
\bibitem{fn:cone} Since the mapping from any $\rho\in\P_s$ to
    a separable CPM via Eq.~\eqref{Eq:State->CPM} is only
    defined up to a positive constant, for the subsequent
    discussion, we might as well consider the cone generated
    by $\P_s$.
\bibitem{fn:F2} Strictly, the inequality \eqref{Eq:F2} is only
    valid when $\lambda_3\neq\lambda_4$. When
    $\lambda_3=\lambda_4$, $\F_2$ degenerates into an edge of
    the polytope $\P_\ld$. If $\ld_1=\ld_2$, $\F_2$ collapses
    into a single point $\vl=\vl_{12}=\vl_{(34)}$.
\bibitem{fn:reverse} By this, of course, we are referring only
    to the states given in Eq.~\eqref{Eq:quasi} that are
    originally of rank three, and which becomes rank two upon
    quasi-distillation. The states given in Eq.~\eqref{Eq:quasi}
    that are of rank two get quasi-distilled to the singlet
    state and so the process is clearly reversible in this
    case.
\end{thebibliography}
\end{document}